\documentclass[reprint,showpacs,preprintnumbers,amsmath,amssymb,prc,floatfix]{revtex4-1}
\usepackage{float}
\usepackage{color}
\usepackage{graphicx}
\usepackage{dcolumn}
\usepackage{bm}
\usepackage{xcolor}
\usepackage{multirow}
\usepackage{threeparttable}
\usepackage{adjustbox}
\usepackage{longtable}
\usepackage{comment}
\usepackage{natbib}
\usepackage{array}
\usepackage[utf8x]{inputenc}
\usepackage[colorlinks,citecolor=blue,linkcolor=red,anchorcolor=blue,filecolor=blue,urlcolor=blue]{hyperref}

\begin{document}
\title{Exploring the giant monopole resonance in superheavy nuclei: A theoretical perspective}

\author{S. Kumar$^1$}
\email{sankm036@gmail.com}
\author{Jeet Amrit Pattnaik$^{2}$}
\email{jeetamritboudh@gmail.com}
\author{S. K. Singh$^{1}$}
\author{R. N. Panda$^{2}$}
\author{M. Bhuyan$^{3}$}
\author{S. K. Patra $^{2}$}
\affiliation{$^1$Department of Physics, Patliputra University, Patna-800020, India}
\affiliation{$^2$Department of Physics, Siksha 'O' Anusandhan Deemed to be University, Bhubaneswar-751030, India}
\affiliation{$^3$Institute of Physics, Sachivalya Marg, Bhubaneswar-751005, India}

\date{\today}

\begin{abstract}
\noindent
Within the relativistic mean field framework, in an extended Thomas-Fermi approximation, we calculate the binding energy and charge distribution radius for the latest superheavy nuclei, synthesised in various laboratories, with atomic numbers $Z = 110-118$. The binding energy and radii are compared with the results obtained from relativistic Hartree calculations along with the experimental data, wherever available, to check the reliability of the methods.  The calculations are extended to estimate the giant monopole resonances to understand the collective vibration of the nucleons for such superheavy nuclei. The giant monopole resonances obtained from scaling calculations are compared with the constraint computations. Furthermore, the results are compared with other known methods, such as the relativistic Random Phase Approximation (RPA) and time-dependent mean field calculations, along with some known lighter nuclei, specifically Zr isotopes (N = 42-86) and O isotopes (N = 10-36). Finally, the nuclear compressibility of the superheavy nuclei is predicted from the energy obtained in the breathing mode.  \\
\end{abstract}
\maketitle
\section{Introduction}
\label{intro} 
\noindent
The superheavy element is one of the most fascinating topics in nuclear physics. So far, up to the atomic number Z = 118 has been synthesised in the laboratory \cite{Og2006}. The heaviest known element $^{294}$Og (Oganesson) has the atomic number Z = 118 and the mass number A = 294 with the neutron number N = 176. The transuranium elements are not found naturally. They are formed artificially in the laboratory and have very short lifetimes with a high radioactive decay rate.  The transuranium elements are useful in many ways, such as nuclear power generation, medical treatment, and scientific research. Because of the short lifetime and difficulties in the synthesis process, the properties of the superheavy elements are not experimentally well known. It is also not clear what the exact atomic weight will be for a particular transuranium element and its position in the Periodic Table. To understand the nature of these superheavy elements in an experimental setup, a proper and thorough theoretical understanding is necessary.  From a theoretical point of view, the non-relativistic \cite{Skyrme,Bohi1979,Chab1998,Butt2019,Kvas2016} and relativistic\cite{Bogu1977,Serot1986,Cent1992,Cent1993,Spei1993,Cent1993a,Lala1997,Cent1998,Patr2001,Patr2002,Patt2021,Patt2023,Patt2024} formalisms are two successful frame-works from which we get the ground state binding energy, charge radii, quadrupole moments and various types of collective oscillations \cite{Bahi2025}. Among the collective properties of nuclei, the monopole and quadrupole vibrations are quite common, and these properties of nuclei give a good amount of information that can be used to study further. Being the heaviest side of the mass table, superheavy nuclei have a large number of protons and neutrons $(\sim 300)$, possess a perfect collective property. To deal with these collective properties, one needs a framework, i.e., the relativistic or the non-relativistic formalisms. Furthermore, the semi-classical approach, such as Thomas-Fermi treatment with its higher order correction on $\hbar$ or direct Hartree-type approach, is more suitable and time-saving in the process of the calculations \cite{Cent1990}. 

One of the fundamental and crucial properties of Nuclear Matter (NM) is the incompressibility at saturation $K_{\infty}$. Mathematically, $K_{\infty}$ is proportional to the second derivative of energy density \cite{Bogu1977,Anki2021,Patt2024b}, i.e., $K_{\infty}{\propto} \frac{\partial^2 {\cal E}}{\partial\rho^2}|_{\rho=\rho_0}$, where $\cal{E}$ = energy density, $\rho$ is the nuclear density and $\rho_0$ is the nuclear density at saturation of the NM. This incompressibility $K_{\infty}$ plays a pivotal role in deciding the nature of the nuclear Equation of State (EoS). For example, with a stiffer EoS, the Tolman-Oppenheimer-Volkoff (TOV) equation predicts a massive Neutron Star (NS) \cite{Kuma2017}. The EoS also determines the flow rate in the Heavy Ion collision experiment \cite{Dani2002}.  It is to be noted that the incompressibility is not a directly measurable quantity. One can extract $K_{\infty}$ from the giant monopole vibration at the resonance condition from the experimental data of finite nuclei. The giant monopole resonances (GMR) are high-frequency, small-amplitude and collective excitation of atomic nuclei. This is a property of many-body systems, characterised by the excitation energy of the system. Thus, the study of GMR of superheavy nuclei is a perfect application of collective phenomena in the field of Nuclear Structure physics. The isoscalar giant monopole resonance (ISGMR) has been reported recently for the medium and heavy mass nuclei~\cite{Bahi2024}. Over the past several decades, the incompressibility of nuclear matter has garnered considerable interest, and multiple methods have been utilized to determine the $K_{\infty}$ \cite{Blai1980,Bohi1979,Maru1989}. Nevertheless, a clear identification of the incompressibility of nuclear matter has proven to be a challenging task because of the lower sensitivity of giant monopole resonance data to incompressibility.

The relativistic mean field models, along with the random phase approximation (RPA), predict the nuclear matter incompressibility at saturation within a range of $K_{\infty}=270\pm{10}$ MeV \cite{Lala1997,Vret2002}. Similarly, the non-relativistic formalisms with Hartree-Fock (HF) plus RPA give this value of $K_{\infty}$ to $ 210- 220$~MeV, which is less than the relativistic approach; on the other hand, the recent experimental value of $K_{\infty}=240\pm{20}$ MeV~\cite{Colo2006, Colo2008,Colo2018,Patt2022}. This discrepancy in $K_{\infty}$ might have come from the diverse behaviour of the density dependence of the symmetry energy in the relativistic and non-relativistic formalism~\cite{Piek2002}. As mentioned earlier, the $K_{\infty}$ is not a direct experimental observable; rather, it is an outcome of the experimental measurement of the isoscalar giant monopole resonance (ISGMR). This fact demands a correct measurement of the excitation energy of the ISGMR. A step forward in this direction is the recent ISGMR calculation in the context of the coherent density fluctuation model \cite{Gaid2025}. Over the last two decades, this fundamental relation of ISGMR and incompressibility has motivated us to study the excitation energy and incompressibility $K^{A}$ of finite nuclei. The excitation energy of the finite nuclei is a collective property that can be studied more accurately in a larger mass system.

In our study, we have taken the heaviest region, which has been synthesised already, of the nuclear landscape for the study of ISGMR and $K^{A}$. Although so much work has been done on ISGMR calculations for heavier nuclei, the superheavy region is still uncovered. A lot of work can be done by taking the superheavy region to explore the many exciting nuclear phenomena. In the scaling method, one can derive the virial theorem for the RMF model of nuclei treated in the extended Thomas–Fermi approach \cite{Patr2002,Patr2001}. The extended Thomas–Fermi solutions satisfy the stability condition against scaling. We apply this semi-classical technique to study collective properties, such as the excitation energy of the breathing mode in superheavy nuclei with the NL3 relativistic parameter set \cite{Lala1997}. Since the superheavy region is not explored properly and the results of this region are unknown with the present method, we first compare the results for a few lower mass nuclei, such as O and Zr \cite{Bisw2015}. Although the well-known NL3 parameter set is quite old and many more parameter sets in the relativistic mean field formalism have been developed, we have chosen the NL3 set in the present calculations due to its merit. The current observation of a higher solar mass neutron star in the multi-messenger experiment \cite{GW190814} suggests a stiffer EoS and the higher side of nuclear matter incompressibility. These properties are well taken in the NL3 force parameters. Further, the NL3 parametrisation gives ground state binding energies, root mean square radii and nuclear densities in very good agreement with the experimental data for known stable and unstable nuclei for both spherical and deformed nuclei. The celebrity relativistic extended Thomas-Fermi approximation (RETF) \cite{Cent1992,Cent1993,Spei1993,Cent1993a,Cent1998,Bisw2015,Bisw2015a} with scaling and constraint approaches are used in the frame-work of the non-linear $\sigma-\omega$ model~\cite {Bogu1977}, which is the $\hbar^2$ correction to the relativistic Thomas-Fermi (RTF) approximation, where variation of density takes care of mostly on the surface of the nucleus~\cite{Cent1990}. It is to be noted that the RETF formalism is more towards the quantal Hartree approximation. It is also verified that the semi-classical approximation, like the Thomas-Fermi method, is very useful in the calculation of collective properties of nuclei, like the giant monopole resonance (GMR)~\cite{Ring1980,Bark1997,Petk1991}.

This paper is organised as follows: In Sec. \ref{theory}, we have summarised the theoretical formalisms, which are relevant for the present analysis. In Sec. \ref{results}, we discuss the results on binding energy obtained from the RETF formalism. We also compared the results with the Hartree calculations as well as experimental/extrapolated data, wherever available. The charge radius and neutron-skin thickness are compared with Hartree calculations as these experimental values are not available for the superheavy nuclei.  After these comparisons, we have elaborated on the giant monopole resonance and compressibility of smaller known nuclei (O and Zr). Then we extended our discussions for the synthesised superheavy nuclei (Z = 110 to 118). The last Sec. \ref{summary} is devoted to a summary and the concluding remarks.

\section{The Relativistic Mean Field Formalism}
\label{theory}
The virial theorem is obtained from the relativistic mean field \cite{Serot1986} Hamiltonian in the framework of scaling invariance within the relativistic Thomas-Fermi (RTF) and relativistic extended Thomas-Fermi (RETF) approximations \cite{Patr2001,Patr2002,Zhu1991,Cent1993,Cent1992,Spei1998,Cent1998,Cent1993a}. Although scaling and constrained calculations have been previously established, the method introduced by Patra et al.~\cite{Patr2001,Patr2002,Zhu1991,mario10,Patr2002} is a novel approach that differs from other scaling techniques. The detailed formalism of the scaling method can be found in Refs.~\cite{Patr2001,Patr2002,Zhu1991,mario10}. For completeness, we have briefly summarised some key expressions, which are essential for the present calculations. We have utilised the non-linear Lagrangian proposed by Boguta and Bodmer~\cite{Bogu1977} to account for many-body correlations that arise from non-linear terms in the $\sigma$-meson self-interaction~\cite{schiff51,schiff51a,schiff50,fujita57,steven01} for nuclear many-body systems. The compressibility modulus of nuclear matter, $K_{\infty}$, decreases significantly with the inclusion of these terms and comes to the experimental limit. This extension of non-linearity over the linear $\sigma-\omega-$ model of Walecka improves the finite nuclei results throughout the mass table, along with the nuclear matter properties.  Based on this model, a large number of relativistic parameter sets are constructed, such as NL1, NL2, NL-SH, NL3, NL3* and many others. Because of the recent experimental data for finite nuclei as well as for NM and NS, the non-linearity is extended further, and more recent forces like G1, G2, G3, and IOPB-I are designed. Taking into consideration the present problem, we have taken the NL3 force in our calculations.

The Hamiltonian for the relativistic mean field nucleon-meson interacting system  is written as ~\cite{Piek2002,Serot1986,Patr2001,Patr2002,Zhu1991}:
\begin{widetext}
\begin{eqnarray}
{\cal H}&= &\sum_i \varphi_i^{\dagger}
\bigg[ - i \vec{\alpha} \cdot \vec{\nabla}
+\beta m^* + g_{v} V + \frac{1}{2} g_{\rho} R \tau_3 
+\frac{1}{2} e {\cal A} (1+\tau_3) \bigg]
\varphi_i 
+ \frac{1}{2} \left[ (\vec{\nabla}\phi)^2 + m_{s}^2 \phi^2 \right]\nonumber\\
&+&\frac{1}{3} b \phi^3
+ \frac{1}{4} c \phi^4
-\frac{1}{2} \left[ (\vec{\nabla} V)^2 + m_{v}^2 V^2 \right]
- \frac{1}{2} \left[ (\vec{\nabla} R)^2 + m_\rho^2 R^2 \right]
- \frac{1}{2} \left(\vec{\nabla}  {\cal A}\right)^2 
\end{eqnarray}
\end{widetext}
The fields for the  ${\sigma}$, ${\omega}$, ${\rho}-$mesons  and  photon are  ${\phi}$, $V$ and $A$, respectively. The $\tau_3$ is the $3^{rd}$ component of the iso-spin of the ${\rho}-$meson.  The actual mass $m$ of the nucleon appears to be missed due to its oscillations in the mesonic medium and acquires the effective mass $m^*=m-g_s\phi$. The $m_s$, $m_v$ and $m_{\rho}$ are the masses for the ${\sigma}-$, $\omega-$ and ${\rho}-$mesons, respectively. The coupling constants $g_s$, $g_v$, $g_{\rho}$ and $e^2/4{\pi}$=1/137 are correspond to the ${\sigma}$, ${\omega}$, ${\rho}$-mesons and photon couplings with the nucleons, respectively. The $b$ and $c$ are the self-couplings for the non-linear ${\sigma}-\omega$ model for the $\sigma-$meson field.
Using the variational principle, the equations of motion for the nucleons and meson fields can be derived.  In the semi-classical approximation,  in terms of density, the Hamiltonian is given by: 
\begin{eqnarray}
{\cal H}&=&{\cal E}+g_v V {\rho}+g_{\rho}R{\rho}_3+e{\cal A}{\rho}_p+{\cal H}_f,
\end{eqnarray}
with
\begin{eqnarray}
{\cal E}&= &\sum_i \varphi_i^{\dagger}
\bigg[ - i \vec{\alpha} \cdot \vec{\nabla} +
\beta m^*\bigg]\varphi_i.
\end{eqnarray}
Here, ${\cal H}_f$ is the free part of the Hamiltonian,  the total density $\rho$ is the sum of proton $\rho_p$ and neutron $\rho_n$ density distribution.

The ground state meson fields are obtained from the Euler-Lagrange equations $\delta {\cal H}/\delta \rho_q = \mu_q$ ($q=n, p$), which are iteratively solved in a self-consistent method. The pairing correlation plays a minor role in the Thomas-Fermi approximation for the giant resonance as well as in the equilibrium property.  It is shown by Vi\~nas and group \cite{xavier11,baldo13} that for open-shell nuclei, the pairing correlation has a marginal effect on ISGMR energy. Thus, we have ignored the pairing contribution in the present semi-classical approach. In the density functional method, the RMF  Hamiltonian is written as:
\begin{align}
{\cal H} 
&= {\cal E} + \frac{1}{2}g_{s}\phi \rho^{\text{eff}}_{s}
+ \frac{1}{3}b\phi^3 + \frac{1}{4} c \phi^4 \nonumber \\
&\quad + \frac{1}{2} g_{v}  V \rho 
+ \frac{1}{2} g_\rho R \rho_3 
+ \frac{1}{2} e {\cal A} \rho_{p}
\label{eqFN8c}
\end{align}

with
\begin{eqnarray}
{\rho}_s^{eff}&= & g_s{\rho}_s-b{\phi}^2-c{\phi}^3.
\end{eqnarray}
Here, we used the scaling technique to calculate the monopole vibration of the nucleus. The baryon density is scaled with a scaling parameter $\lambda$, which is defined in the normalised form \cite{Patr2001,Patr2002,Zhu1991} as:
\begin {eqnarray}
{\rho}_{\lambda}\left(\bf r \right)&=&{\lambda}^3{\rho}\left(\lambda r \right),
\end{eqnarray}
where ${\lambda}$ is a collective coordinate associated with the monopole vibration. 
Since the  Fermi momentum $K_{Fq}$ and density are related to each other, the scaled Fermi momentum $K_{Fq\lambda}$ is given by
\begin {eqnarray}
K_Fq{\lambda}&=&\left[3{\pi}^2{\rho}_q
\lambda\left(\bf r \right)\right]^{\frac{1}{3}}.
\end {eqnarray}
Similarly, the $\phi$, $V$, $R$ and Coulomb fields are scaled. Because the source term $\phi$ field contains the $\phi$ field itself, it can not be scaled similarly to the density and momentum.  In the present approach (semi-classical formalism), the energy and density are scaled as 
\begin{eqnarray}
{\cal E_{\lambda}}(\bf r)&=&{\lambda}^4 {\tilde{\cal E}}(\lambda \bf r)
= \lambda^4[\tilde{\cal E}_{0}(\lambda \bf r)+
\tilde{\cal E}_2(\lambda \bf r)],
\end {eqnarray}
\begin{eqnarray}
\rho_{{s\lambda}}(\bf r)={\lambda^3}{\tilde{\rho}}_{s}{(\lambda{\bf r})}.
\end{eqnarray}
The $\sim$ is denoted as an implicit dependence of $ \tilde m^* $. Now, using the scaled variables, the scaled semi-classical Hamiltonian can be written as:
\begin{align}
{\cal{H}}_{\lambda} & = & {\lambda^3}{\lambda}{\tilde{\cal{E}}}+\frac{1}{2}{g_s}
\phi_{\lambda}{{\tilde{\rho}}_s^{eff}}\nonumber
+\frac{1}{3}\frac{b}{\lambda^3}\phi_\lambda^3
+\frac{1}{4}\frac{c}{\lambda^3}{\phi_\lambda}^4
+\frac{1}{2}{g_v}V_\lambda{\rho}
\nonumber \\[3mm]
 & &  
+\frac{1}{2}g_{\rho}R_\lambda {\rho_3}+\frac{1}{2}e{A}_\lambda{\rho}_p
\end{align}
 The scaled monopole excitation energy ${E}_{S}$ is defined as
${E}_{S}={\sqrt{\frac{C_m}{B_m}}}$, ${C_m}$=the restoring force and
$B_m$= the mass parameter. The restoring force is defined as\cite{Patr2001,Patr2002,Zhu1991} $C_m=\frac{\partial^2{\cal E_{\lambda}}}{\partial{\lambda^2}}|_{\lambda=1}$, which after operation is obtained as:
\begin{align}
C_m = \int dr\, \bigg[
& -m \frac{\partial \tilde{\rho}_s}{\partial \lambda}
+ 3\left(m_s^2 \phi^2 + \frac{1}{3} b \phi^3 
- m_v^2 V^2 - m_{\rho}^2 R^2 \right) \nonumber \\
& - \left(2 m_s^2 \phi + b \phi^2 \right) 
\frac{\partial \phi_\lambda}{\partial \lambda}
+ 2 m_v^2 V \frac{\partial V_\lambda}{\partial \lambda}
+ 2 m_{\rho}^2 R \frac{\partial R_\lambda}{\partial \lambda}
\bigg]_{\lambda=1}
\label{eq:Cm}
\end{align}

Similarly, the mass parameter of the monopole vibration is expressed as $B_m=\frac{\partial^2{\cal E_{\lambda}}}{\partial{\dot{\lambda^2}}}|_{\dot{\lambda=1}}$, (i.e., the double derivative of the scaled energy with the collective velocity $\dot{\lambda}$)
as:
\begin{eqnarray}
B_{m}=\int{dr}{U(\bf r)}^2{\cal {H}},
\end {eqnarray}
where $U(\bf r)$ is the displacement field, which is obtained from the relation
between collective velocity $\dot{\lambda}$ and velocity of the moving frame,
\begin {eqnarray}
U(\bf r)=\frac{1}{\rho(\bf r){\bf r}^2}\int{dr'}{\rho}_T(r'){r'}^2,
\end {eqnarray}
where, ${\rho}_T $ is the transition density defined as
\begin {eqnarray}
{{\rho}_T(\bf r)}=\frac{\partial{\rho_\lambda(\bf r)}}{\partial{\lambda}}\bigg|_{\lambda = 1}
=3 {\rho}(\bf r)+r \frac{\partial{\rho(\bf r)}}{\partial r}.
\end {eqnarray}
On substituting the value of the transition density $\rho_T(\bf{r})$ in Eq. (13) and after the simplification of the integral, the displacement field becomes $U(\bf r)=r$. 
Ultimately, the mass parameter is written as
$B_m=\int{dr}{r}^2{\cal H}$.
In non-relativistic limit, ${B_m}^{nr}=\int{dr}{r^2}m{\rho}$. The scaled excitation energy can be expressed by different moments $m_k=\int_{0}^{\infty}E^kS(E)dE$, where S(E) is the strength function and $k$ is an integer. The average excitation energy with the scaling approach is ${E_S}$=$\sqrt{\frac{m_3}{m_1}}$. The expressions for ${m_3}$ and ${m_1}$ can be found in Refs. \cite {Bohi1979,Cent2005}. Apart from the scaling calculation, the monopole vibration can also be studied with a constrained approach \cite{Bohi1979,Maru1989,Boer1991,Stoi1994,Stoi1994a}. In this method, one has to solve the
constrained functional equation \cite{Patr2001,Patr2002}:
\begin{eqnarray}
\int{dr}\left[{\cal H}-{\eta}{r}^2 {\rho}\right]=E(\eta)-\eta\int{dr}{r}^2\rho.
\end{eqnarray}
The constrained is ${\langle {R^2}\rangle}_0 ={\langle{r^2}\rangle}_m$.
The constrained energy $E(\eta)$ can be expanded in a harmonic approximation as
\begin{eqnarray}
E(\eta)& = &E(0)+\frac{\partial E(\eta)}{\partial \eta}\big|_{\eta =0}
+\frac{\partial^2{E(\eta)}}{\partial{\eta}^2}|_{\eta =0}.
\end {eqnarray}
The second-order derivative in the expansion is related to the constrained incompressibility ${K}^A$ for a finite nucleus with mass number $A$ as
\begin{eqnarray}
{K_C}^{A} = \frac{1}{A} {R_0}^{2} \frac{\partial^2{E \eta}}
{\partial{R_\eta}},
\end{eqnarray}
and the constrained monopole excitation energy ${E_C}$ as
\begin{eqnarray}
{{E_C}={\sqrt{\frac{A {K_C^A}}{B_C^m}}}}.
\end{eqnarray}
Similar to the scaled excitation energy, the average resonance energy for the constraint approach can be written with the sum-rule notation as ${E_C}$=$\sqrt{\frac{m_1}{m_{-1}}}$. The scaled and constraint energies are the upper and lower bounds of the resonance, respectively, and the average resonance width $\Sigma$ is their difference, which is defined as \cite{Bohi1979,Cent2005} $\Sigma=\frac{1}{2}{\sqrt{E_{S}^2-E_{C}^2}}$.

\section{RESULTS AND DISCUSSIONS} \label{results}
In this section, we discuss the obtained results of binding energy, root mean square radius and nucleon density distribution for some selected nuclei. The transition density is also given to understand the change in nucleon distribution inside the superheavy nuclei in the monopole vibrations. The theoretical prediction of the monopole excitation energy in the scaling approach $E_S$ along with the constraint calculation $E_C$ is discussed in detail in the following subsections through Tables (1-2) and Figs. (1-9). 
\subsection{Binding energy}
\begin{figure}
\includegraphics[width=1.0\columnwidth]{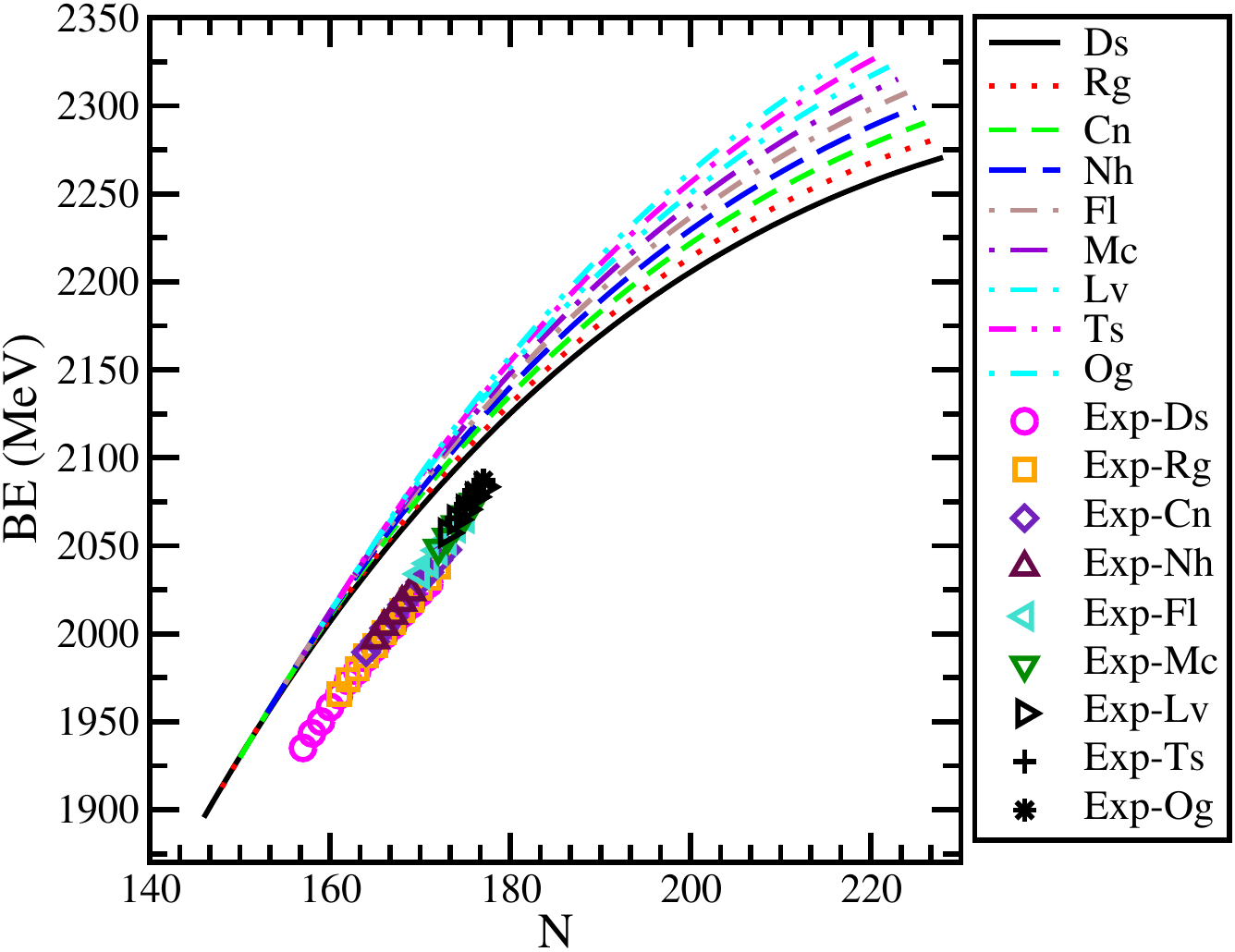}
\caption{The binding energy for superheavy nuclei $Z = 110-118$ isotopes compare with available experimental/extrapolated  data~\cite{Wang2017}}. 
\label{fig1}
\end{figure}

\begin{table*}[t]
\caption {The binding energy (BE) obtained from the relativistic Extended Thomas-Fermi (RETF) calculations for some selected nuclei is compared with the relativistic Hartree approximation (RHA) as well as with the experimental data \cite{Wang2017}, wherever available, for the superheavy nuclei with atomic number $Z=110-118$. The energy is in MeV, and calculations are done with NL3 parameters set.}
\renewcommand{\tabcolsep}{0.10cm}
\renewcommand{\arraystretch}{1.0}
\begin{tabular}{cccccccc}
\hline\hline
Nucleus & ETF (BE) & RHA (BE) & Expt. (BE)& Nucleus & ETF (BE) & RHA (BE) & Expt. (BE)\\
\hline
$^{267}$Ds & 1985.3 &1929.4  & 1935.2&
$^{268}$Ds & 1992.7 &1937.0 & 1943.5\\
$^{269}$Ds & 1999.97 &1944.7  & 1950.3 &
$^{270}$Ds  & 2007.0 &1952.3  & 1958.5 \\
$^{271}$Ds  & 2014.0 &1959.8 & 1965.3 &
$^{272}$Ds & 2020.8 &1967.3   & 1973.4 \\
$^{273}$Ds & 2027.5 &1974.8   & 1979.0 &
$^{274}$Ds & 2034.2 &1982.2    & 1986.2 \\
$^{275}$Ds & 2040.7 &1989.0    & 1992.1 &
$^{276}$Ds & 2047.1 &1995.7    & 1999.1 \\
$^{277}$Ds & 2053.3 &2002.5    & 2004.7 &
$^{278}$Ds & 2059.5 &2009.2    & 2011.3 \\
$^{279}$Ds & 2065.5 &2015.9   & 2016.6 &
$^{280}$Ds & 2071.5 &2022.5    & 2023.4\ \\
$^{281}$Ds & 2077.3 &2029.1    & 2028.5 \\
\hline
$^{272}$Rg & 2016.0 & 1961.9 & 1965.7&
$^{273}$Rg & 2023.1 & 1969.6 & 1974.1 \\
$^{274}$Rg & 2030.0 & 1977.3 & 1980.2&
$^{275}$Rg & 2036.8 & 1984.9 & 1987.4 \\
$^{276}$Rg & 2043.5 & 1991.9 & 1993.3 &
$^{277}$Rg & 2050.0 & 1998.9 & 2000.5 \\
$^{278}$Rg & 2056.5 & 2005.8 & 2006.6 &
$^{279}$Rg & 2062.8 & 2012.7 & 2013.3 \\
$^{280}$Rg & 2069.1 & 2019.5 & 2019.4 &
$^{281}$Rg & 2075.2 & 2026.3 & 2026.0 \\
$^{282}$Rg & 2081.2 & 2033.1 & 2031.5 &
$^{283}$Rg & 2087.1 & 2039.8 & 2038.2 \\
\hline
$^{276}$Cn & 2039.0& 1987.5 & 1989.7 &
$^{277}$Cn & 2045.9 & 1994.7 & 1995.8 \\
$^{278}$Cn & 2052.6 & 2001.8 & 2003.3 &
$^{279}$Cn & 2059.3 & 2008.9 & 2009.4 \\
$^{280}$Cn & 2065.8 & 2016.0 & 2016.6 &
$^{281}$Cn & 2072.2 & 2022.9 & 2022.4 \\
$^{282}$Cn & 2078.5 & 2029.9 & 2029.6 &
$^{283}$Cn & 2084.7 & 2036.8 & 2035.1 \\
$^{284}$Cn & 2090.7 & 2043.7 & 2042.0 &
$^{285}$Cn & 2096.7 & 2049.5 & 2047.4 \\
\hline
$^{278}$Nh & 2047.9 & 1997.2 & 1996.6 &
$^{279}$Nh & 2054.8 & 2004.5 & 2004.3 \\
$^{280}$Nh & 2061.6 & 2011.7 & 2010.4 &
$^{281}$Nh & 2068.3 & 2018.9 & 2017.9 \\
$^{282}$Nh & 2074.9 & 2026.1 & 2023.9 \\
\hline
$^{284}$Fl & 2083.9 & 2036.4  & 2034.0 &
$^{285}$Fl & 2090.5& 2043.7 & 2040.0 \\
$^{286}$Fl &2096.9 & 2051.0  & 2047.5 &
$^{287}$Fl &2103.2 &  2057.0 & 2053.2 \\
$^{288}$Fl & 2109.4 & 2063.0 & 2060.4 &
$^{289}$Fl & 2115.5 & 2069.0 & 2065.8 \\
\hline
$^{287}$Mc & 2099.4 & 2053.6 & 2048.6 &
$^{288}$Mc & 2105.9 & 2059.7 & 2054.9 \\
$^{289}$Mc & 2112.3 & 2065.8 & 2062.0 &
$^{290}$Mc & 2118.5 & 2071.9 & 2067.7 \\
$^{291}$Mc & 2124.7 & 2078.0 & 2074.8 \\
\hline
$^{289}$Lv & 2108.2 & 2062.1 & 2057.4&
$^{290}$Lv & 2114.7 & 2068.3 & 2064.8 \\
$^{291}$Lv & 2121.3 & 2074.5 & 2070.8 &
$^{292}$Lv & 2127.5 & 2080.7 & 2078.2 \\
$^{293}$Lv & 2133.7 & 2086.8 & 2083.5 \\
\hline
$^{291}$Ts & 2116.8 & 2070.6 & 2065.5 &
$^{292}$Ts & 2123.4 & 2076.9 & 2071.7 \\
$^{293}$Ts & 2129.9 & 2083.2 & 2078.8 &
$^{294}$Ts & 2136.3 & 2089.432 & 2085.0 \\
\hline
$^{293}$Og & 2125.3 & 2079.1 & 2073.6 &
$^{294}$Og & 2132.0 & 2085.4 & 2081.2 \\
$^{295}$Og & 2138.6 & 2091.8 & 2087.1 \\
\hline \hline
\end{tabular}
\label{tab1}
\end{table*}
\begin{table*}
\caption {The charge radius $R_{ch}$ and neutron-skin thickness $\triangle{r}=R_n-R_p$ obtained from the relativistic extended Thomas-Fermi (RETF) and relativistic Hartree approximation for some selected superheavy nuclei with atomic number $Z = 110-118$. The radii are in fm.}
\renewcommand{\tabcolsep}{0.1cm}
\renewcommand{\arraystretch}{1}
\begin{tabular}{cccccccccc}
\hline\hline
Nucleus & ETF ($R_{\rm ch}$) & RHA ($R_{\rm ch}$) &
ETF($\triangle{r}$) & RHA ($\triangle {r}$)  & Nucleus & ETF ($R_{\rm ch}$) & RHA ($R_{\rm ch}$) &
ETF($\triangle{r}$) & RHA ($\triangle{r}$) \\
\hline
$^{267}$Ds &6.076 & 6.114 &0.137 & 0.144 & $^{268}$Ds &6.083 & 6.118 &0.142 & 0.150 \\
$^{269}$Ds &6.089 & 6.123 &0.147 & 0.155 & $^{270}$Ds &6.096 & 6.128 &0.152 & 0.161 \\
$^{271}$Ds &6.102 & 6.132 &0.156& 0.166 & $^{272}$Ds &6.109& 6.136 &0.161 & 0.171 \\
$^{273}$Ds &6.115 & 6.141 &0.166 & 0.176 & $^{274}$Ds &6.122& 6.145 &0.170 & 0.181 \\
$^{275}$Ds &6.128 & 6.152 &0.175 & 0.188 & $^{276}$Ds &6.135 & 6.159 &0.180 & 0.194 \\
$^{277}$Ds &6.142 & 6.166 &0.184 & 0.200 & $^{278}$Ds &6.148 & 6.173 &0.189 & 0.206 \\
$^{279}$Ds &6.155& 6.180 &0.194 & 0.212 & $^{280}$Ds &6.161 & 6.186 &0.198 & 0.217 \\
$^{281}$Ds &6.168 & 6.193 &0.203 & 0.222 \\
\hline
$^{272}$Rg &6.113 & 6.139 &0.148 & 0.160 & $^{273}$Rg &6.119 & 6.143 &0.153 & 0.165 \\
$^{274}$Rg &6.126 & 6.148 &0.158 & 0.170 & $^{275}$Rg &6.132 & 6.152 &0.162 & 0.175 \\
$^{276}$Rg &6.139 & 6.159 &0.167& 0.181 & $^{277}$Rg &6.145 & 6.166 &0.172 & 0.187 \\
$^{278}$Rg &6.152 & 6.173 &0.176 & 0.193 & $^{279}$Rg &6.158 & 6.180 &0.181 & 0.199 \\
$^{280}$Rg &6.165 & 6.187 &0.186& 0.198 & $^{281}$Rg &6.171 & 6.194 &0.190 & 0.203 \\
$^{282}$Rg &6.178 & 6.201 &0.195 & 0.203 & $^{283}$Rg &6.185 & 6.208 &0.200 & 0.208 \\
\hline
$^{276}$Cn &6.143 & 6.158 &0.155& 0.169 & $^{277}$Cn &6.149 & 6.166 &0.159 & 0.175 \\
$^{278}$Cn &6.156& 6.173 &0.164 & 0.181 & $^{279}$Cn &6.162 & 6.180 &0.168 & 0.187 \\
$^{280}$Cn &6.169 & 6.187 &0.173 & 0.193 & $^{281}$Cn &6.175 & 6.194 &0.178 & 0.199 \\
$^{282}$Cn &6.182 & 6.201 &0.182 & 0.202 & $^{283}$Cn &6.188 & 6.208 &0.187 & 0.208 \\
$^{284}$Cn &6.195 & 6.215 & 0.191& 0.212 & $^{285}$Cn &6.201 & 6.217 &0.196& 0.222 \\
\hline
$^{278}$Nh &6.159 & 6.158 &0.152& 0.169 & $^{279}$Nh &6.166 & 6.166 &0.156 & 0.175 \\
$^{280}$Nh &6.172& 6.173 &0.161 & 0.180 & $^{281}$Nh &6.179 & 6.180 &0.165 & 0.185 \\
$^{282}$Nh &6.185 & 6.187 &0.170 & 0.191 \\
\hline
$^{284}$Fl &6.202 & 6.215 &0.167 & 0.189 & $^{285}$Fl &6.208 & 6.222 &0.171 & 0.194 \\
$^{286}$Fl &6.215 & 6.229 &0.176 & 0.198 & $^{287}$Fl &6.221 & 6.231 &0.180 & 0.207 \\
$^{288}$Fl &6.228 & 6.233 &0.185 & 0.216 & $^{289}$Fl &6.234& 6.236 &0.189 & 0.225 \\
\hline
$^{287}$Mc &6.225 & 6.238 &0.168& 0.192 & $^{288}$Mc &6.231 & 6.240 &0.172 & 0.201 \\
$^{289}$Mc &6.238& 6.242 &0.177 & 0.209 & $^{290}$Mc &6.244 & 6.245 &0.181& 0.218 \\
$^{291}$Mc &6.251 & 6.247 &0.186 & 0.226 & & & & & \\
\hline
$^{289}$Lv &6.241 & 6.249 &0.165 & 0.194 & $^{290}$Lv &6.248 & 6.251 &0.169 & 0.202 \\
$^{291}$Lv &6.254 & 6.254 &0.174 & 0.211 & $^{292}$Lv &6.261 & 6.256 &0.178 & 0.219 \\
$^{293}$Lv &6.267 & 6.259 &0.182 & 0.227 & & & & & \\
\hline
$^{291}$Ts &6.258 & 6.260 &0.162 & 0.196 & $^{292}$Ts &6.264 & 6.263 &0.170 & 0.204 \\
$^{293}$Ts &6.270 & 6.265 &0.170 & 0.212 & $^{294}$Ts &6.277 & 6.268 &0.175 & 0.220 \\
\hline
$^{293}$Og &6.274 & 6.271 &0.158 & 0.197 & $^{294}$Og & 6.280& 6.274 &0.163 & 0.205 \\
$^{295}$Og &6.287 & 6.277 &0.167 & 0.213 & & & & & \\
\hline\hline
\end{tabular}
\label{tab2}
\end{table*}
The binding energy (BE) of finite nuclei is a crucial quantity in the study of nuclear physics. This is the most accurately measured quantity among the known nuclear observables. Thus, knowing the calculated binding energy and how far the method can reproduce is quite instructive. The BE values for some known nuclei in the superheavy region are displayed in Table \ref{tab1}.  For few representative cases, starting from $^{267}$Ds to $^{295}$Og, with neutron number $N=157-177$. 
The binding energy in an isotopic chain gives an idea of the energy required to separate a neutron and also acts as a symbol of nuclear stability. Here, we discuss the binding energy results of our calculation for superheavy nuclei $Z = 110-118$ using the NL3 parameter set in the framework of RETF. The semi-classical results are compared with the relativistic Hartree calculations \cite{Lala1997} along with the experimental data \cite{Wang2017}. The results are depicted in Table \ref{tab1} and Fig. \ref{fig1}. From the table, one can find that, in general, we get a constant overbinding of $\sim 50$ MeV for almost all the isotopes considered in the mass region in RETF. In other words, we get about $2\%$ more binding with the RETF formalism than the experimental data. The binding energy difference between the experimental and RETF approach can be bridged out with a scaling factor $f_s \sim 1.024$, i.e., if we divide the RETF results of binding energy by a factor $f_s$, then we will get the exact experimental binding energy. This discrepancy in bindings between RETF and data may be due to the exclusion of the shell structure of the nucleus in the calculation. However, the quantal RMF binding energy is quite close to the experimental data.  The RMF or Hartree calculation under-predicts the RETF results. 

The RMF results of binding energy are about 4 MeV, which differs from the data, which is almost equal to the experimental observation.  The binding energy calculations towards the neutron dripline in an isotopic chain increase monotonously with neutron number, and it becomes stiffer and stiffer with atomic number. Due to the difficulties in the mass measurement and isotopes synthesis, it is difficult to obtain the binding energy data for a wide range of isotopes, unlike the light or medium mass region of the mass table; hence, for a few cases, the experimental extrapolated data are used in the comparison. In general, although the RETF calculation is a semi-classical approach, its prediction about the binding energy is quite close to the experimental results, which can be seen from the table and figure (Table 1 and Figure 1) \cite{Wang2017}.
\subsection{The charge radius and neutron-skin thickness $R_n-R_p=\triangle{r}$ of superheavy nuclei with $Z = 110-118$}
\begin{figure}
\includegraphics[width=1.05\columnwidth]{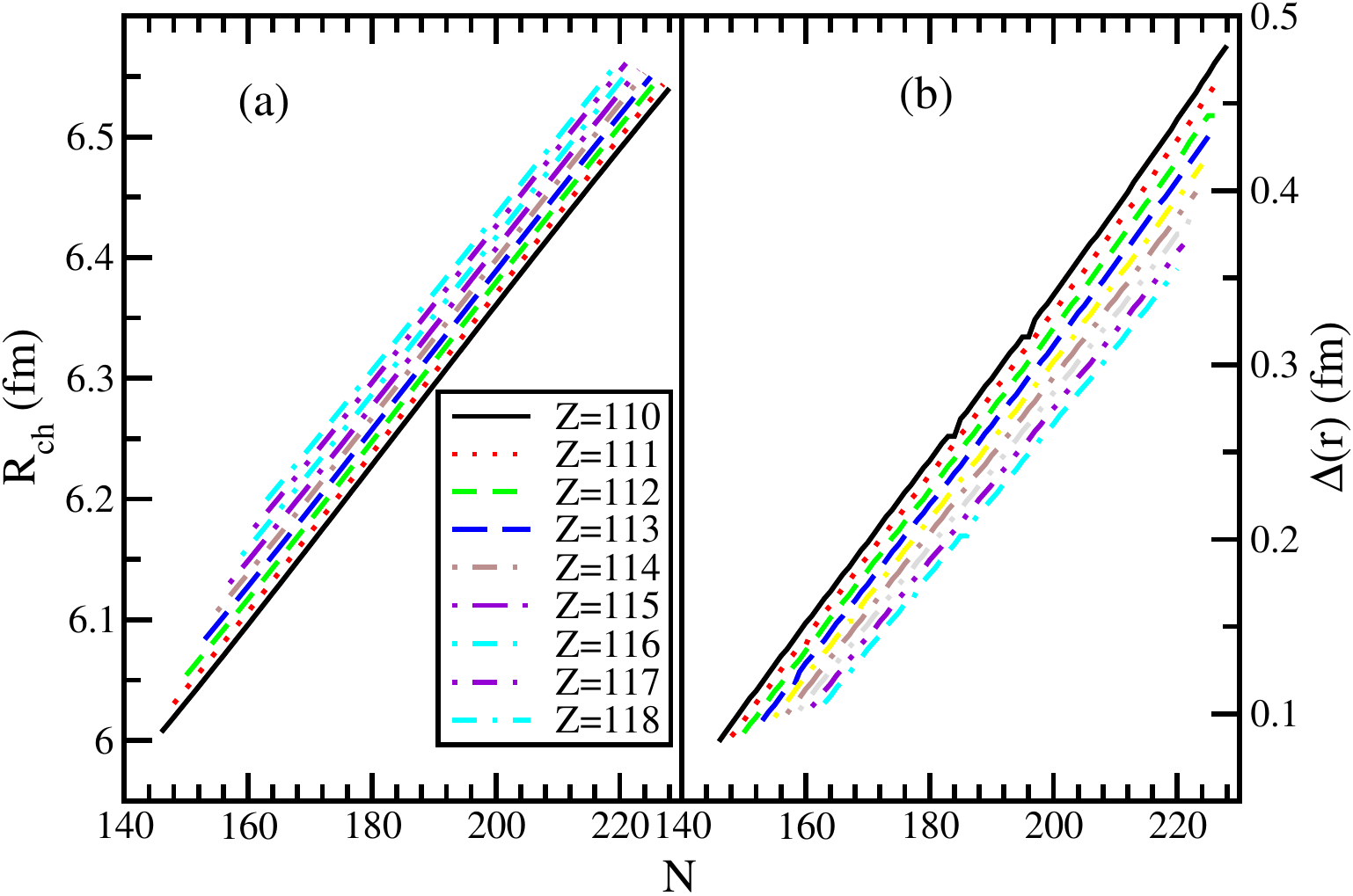}
\caption{The charge radius $R_{ch}$ and neutron-skin thickness $\triangle{r}=R_n-R_p$ for superheavy element with atomic number  $Z=110-118$ using NL3 set. }
\label{fig2}
\end{figure}

The nuclear charge radius $R_{ch}$ is one of the most useful observables that provides information on how effective interactions affect the nuclear structure and clarify the nuclear shell model. In panel (a) of Fig. \ref{fig2}, we examine the findings of charge radius $R_{ch}$ for superheavy nuclei with atomic numbers $Z=110-118$. The $R_{ch}$ is determined by employing the NL3 parameters within the context of relativistic extended Thomas-Fermi formalism. We have also depicted the relativistic Hartree (RHA) results of $R_{ch}$ and neutron-skin thickness ($\triangle{r}=R_n-R_p$) for comparison in Table \ref{tab2}.  The charge radius rises linearly with the neutron count in an isotopic series. In panel (b) of Fig. \ref{fig2}, we present the neutron skin thickness $\Delta{r}=R_n-R_p$ on femtometers (fm) as a function of the neutron number for the heavy nuclei $Z=110-118$. The thickness of the skin increases consistently in magnitude with a rise in the neutron count for every isotope of superheavy nuclei across all elements. This slow rise in the $\triangle{r}$ can be explained by the rearrangement of nucleons due to nuclear interactions when an additional neutron is added while maintaining the atomic number (Z). Since we are analysing the RETF result, which is a semi-classical approximation, the detailed structure of the nucleus is missing. However, the internal structure of the nucleus is visible in the RHA calculations, which can be seen from the $R_{ch}$ and $\Delta{r}$ in Table \ref{tab2}. On average, the RETF and RHA are similar and can be understood as the applicability of the RETF formalism to study the collective properties of the nucleus, including the giant monopole resonances.

\subsection{The nuclear and transition density}
\begin{figure}
\includegraphics[width=1.05\columnwidth]{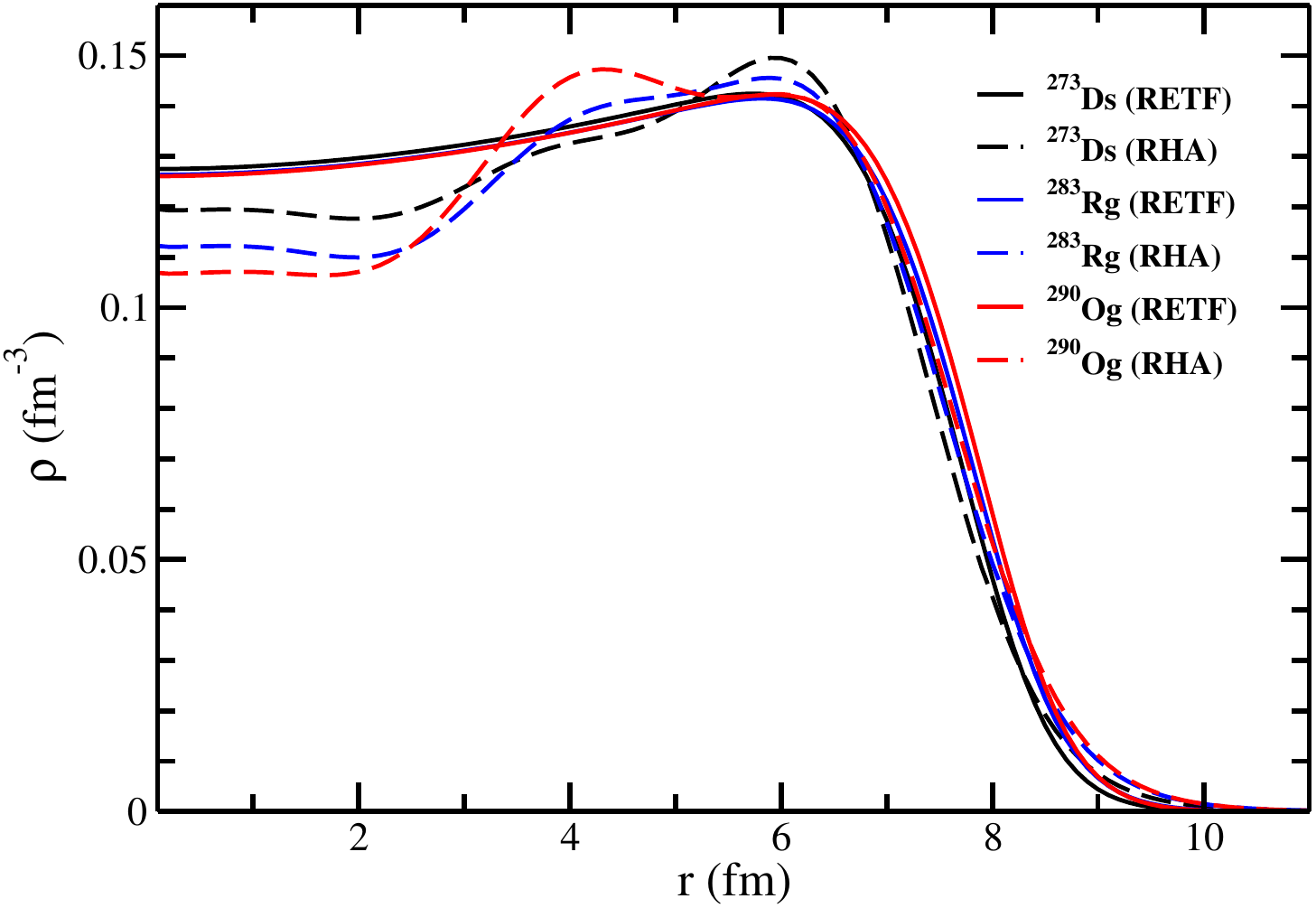}
\caption{The total relativistic extended Thomas-Fermi nuclear densities are compared with the relativistic Hartree approximation (RHA) for a few selected nuclei.}
\label{fig3}
\end{figure}

In this subsection, we analyse the nuclear density and subsequently the transition density for some selected nuclei in the superheavy region. The nuclear density plays a pivotal role not only in the ground state but also in the collective excited state. Thus, it is a crucial quantity for determining the collective oscillations of heavy nuclei.  The RETF density and the quantal RHA density are compared in Fig. \ref{fig3}. We find that both densities are comparable to each other, except for the disappearance of shell structure in the RETF result. This structure of the density is expected in a semi-classical calculation. On the other hand, the internal shell structure is visible in the RHA density. In general, the RETF density is good enough for our subsequent collective calculations.  
\begin{figure}
\includegraphics[width=1.0\columnwidth]{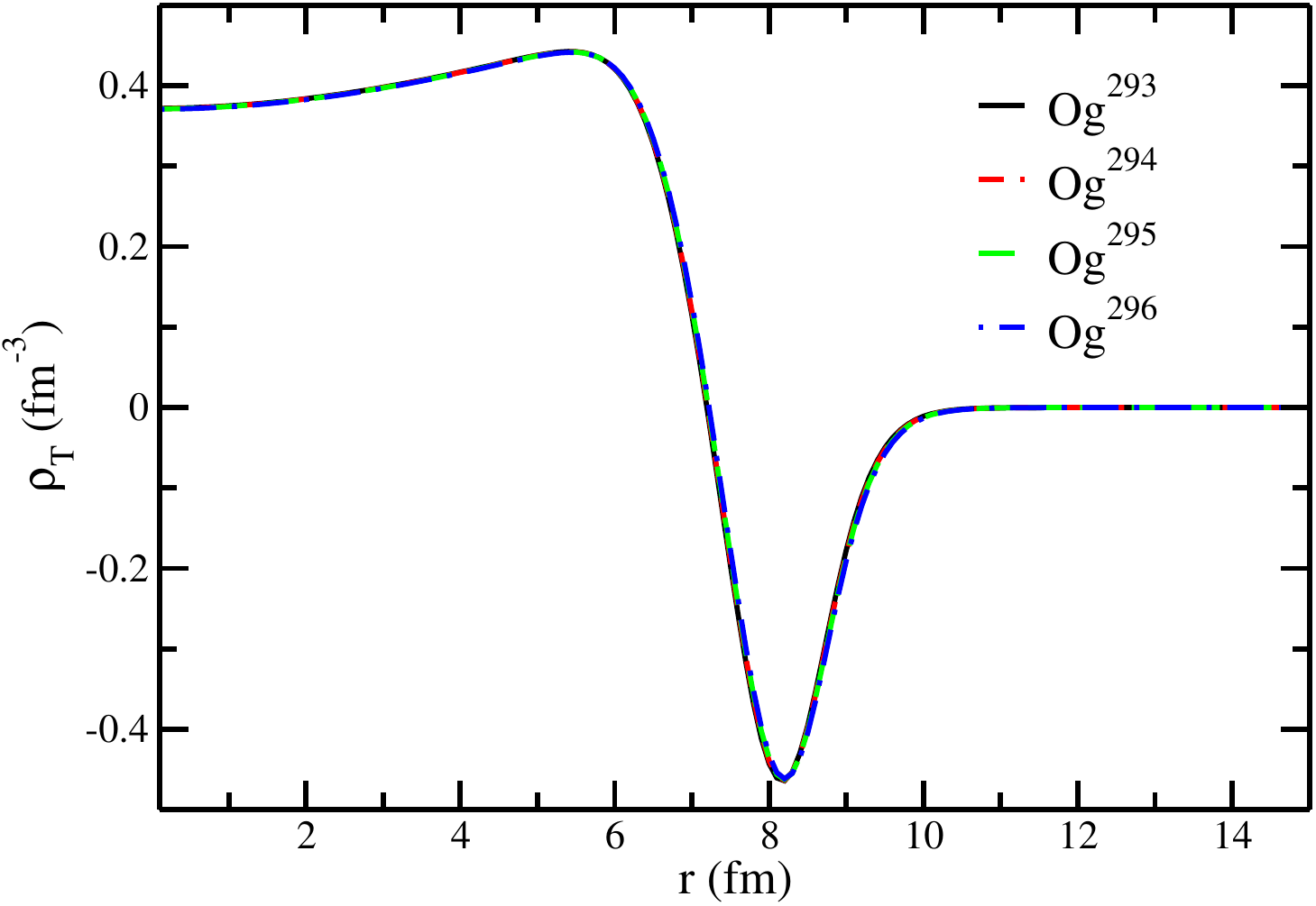}
\caption{The transition density $\rho_T$ for $^{293,294,295,296}$Og with RETF method using NL3 force.}
\label{fig4}
\end{figure}
After knowing the applicability of the RETF density for the collective oscillation, we computed the transition density for some nuclei. The transition density is interpreted in the excitation of the collective mode of vibration, i.e. the movement of the nucleus as a whole. In this case, single-particle nucleons change their energy level.  In the present case, the rate of change of nuclear density with respect to the scaling parameter $\lambda$ at $\lambda = 1$ [see eq. (14)]. The RETF results for $^{293-296}$Og are shown in Fig. \ref{fig4}. This figure helps us to understand the structural evolution during a collective transition.   
\subsection{Giant monopole energy and finite nuclear incompressibility of light mass region}
\begin{figure}
\includegraphics[width=01.0\columnwidth]{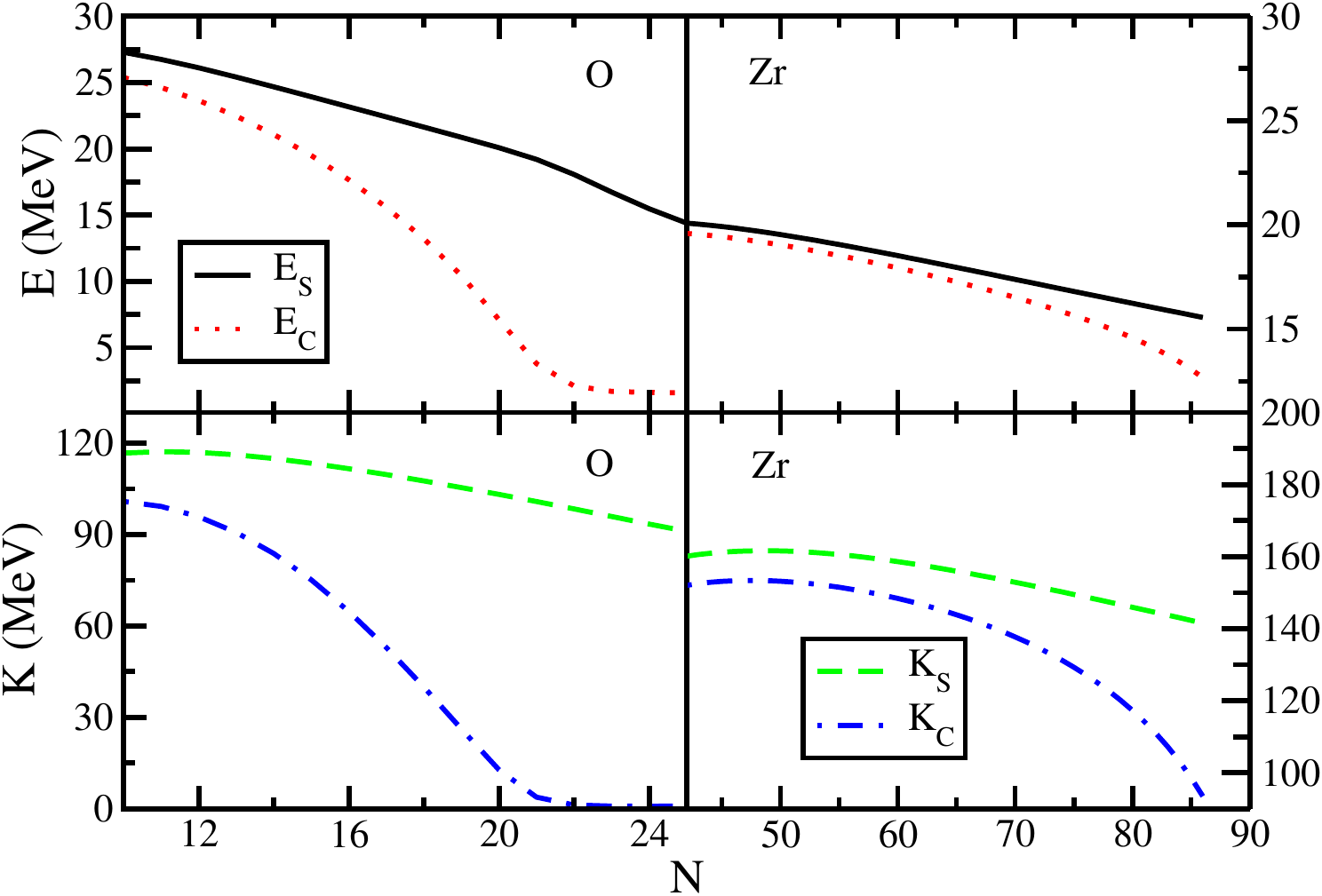}
\caption{(Color online) The giant monopole resonance excitation energy and incompressibility modulus for O and Zr in scaling and constraint method.}
\label{fig5}
\end{figure}

Before going to investigate the isoscalar giant monopole resonances (ISGMR) for the superheavy nuclei, we reproduced the results of our earlier calculations for the O and Zr isotopes. The calculated results are shown in Fig. \ref{fig5} as a function of neutron numbers. The results for the constraint calculations are also shown in the same figure. The sum rule weighted factor expressed that the monopole scaled energy $E_S$ is the ratio of the $m_3$ to the $m_1$, while the monopole constrained energy $E_C$ is the ratio of the $m_3$ to $m_2$. Thus, we obtain the higher side of the monopole resonance spectrum in a scaling calculation of the monopole excitation energy. However, the constraint calculation gives the lower side of the spectrum. The difference between the scaling and constraint excitation energy estimates the width of the spectrum, which is defined as $\Sigma=\frac{1}{2}\sqrt{E_S^2-E_C^2}$. The results for the O isotopes extended from $N=10-26$  and those of the Zr isotopes from $ N=42–86$ as shown in Fig. \ref{fig5}. The $E_S$ is denoted by the solid line and $E_C$ is by the dotted line. It is always found that $E_C\leq E_S$ and both the excitation energies decrease with an increase in neutron number in an isotopic series, but the $\Sigma$ value increases with neutron number and it is maximum near the dripline. The obtained values of excitation energies for $^{16}$O from the scaling and constrained calculations are 27.83 MeV and 25.97 MeV, respectively, which are above the range of experimental results $21.13\pm0.49$ MeV~\cite{Youn2013}. Similarly to the case of $^{16}$O, the predicted excitation energy $E_S=19.53$ MeV and $E_C=19.03$ MeV for $^{90}Zr$ with the experimental value of $E_{expt}=17.89\pm0.20~ $MeV~\cite{Youn2013}. The theoretical results on the monopole excitation energy give the impression that the RETF approximation is more suitable for large systems with a large number of nucleons and could be more appropriate for the superheavy region.

The scaling and constraint incompressibility $K^A_{S}$ and $K^A_C$ for oxygen isotopes starting with $N=10$ to $N=26$ are given in the lower panel of Fig. \ref{fig5}. Similarly, the $K^A$ for scaling and constraint values are also shown in the same figure for Zirconium isotopes.
The $K^A$ results for both the Oxygen and Zirconium isotopes monotonously decrease with the increase in neutron number.  The rate of decrease in constraint $K^A_C$ is much faster than that of $K^A_{s}$. It is almost zero at N = 21, showing the incompressible nature of Oxygen near the dripline. The general trend in the behaviour of excitation energy and incompressibility is almost the same, with a change in the energy scale. In case of the Zr-isotopes, the difference in scaling and constraint incompressibility $\Delta K=K_S-K_C$ with the neutron number N, we find the increasing trend of $\Delta K$. 

\subsection{Giant monopole energy and nuclear incompressibility of superheavy nuclei with atomic number Z = 110-118}
\begin{figure}
\includegraphics[width=01.05\columnwidth]{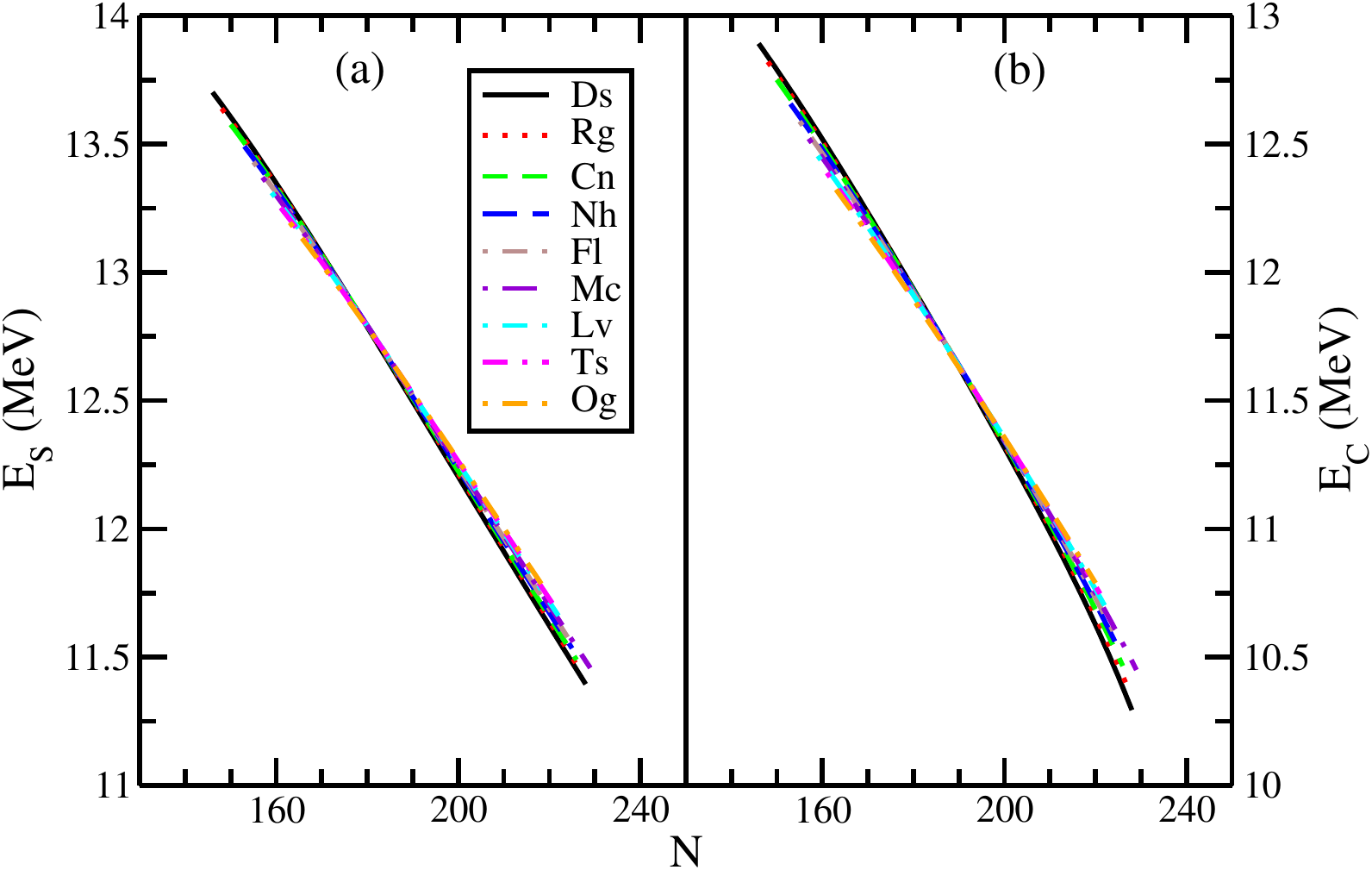}
\caption{The giant monopole resonance excitation energy for superheavy nuclei $Z=110-118$ in scaling and constraint methods as a function of neutron number N. (a) is for the scaling calculation and (b) is for the constraint case.}
\label{fig6}
\end{figure}
\begin{figure}
\includegraphics[width=1.1\columnwidth]{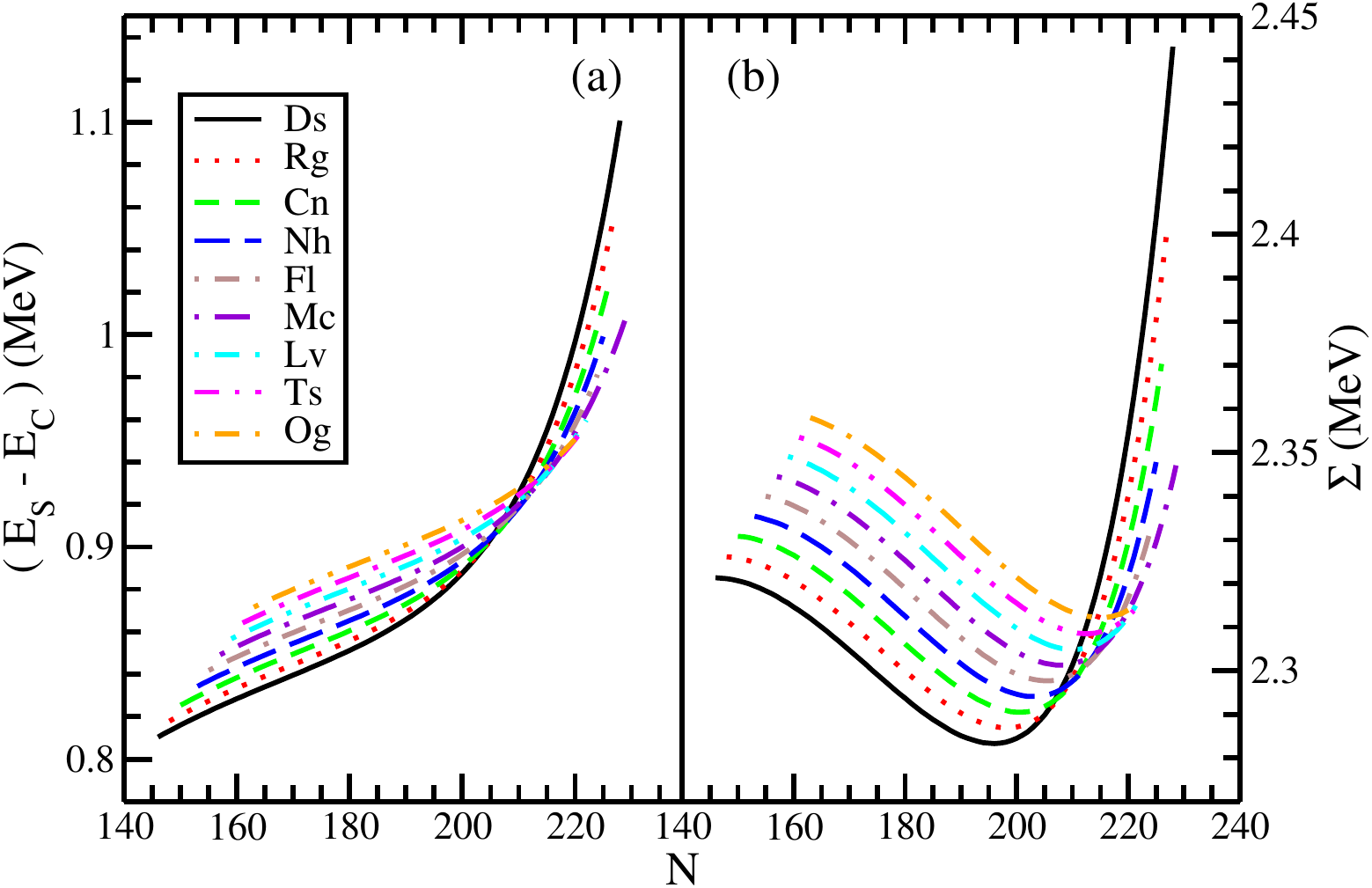}
\caption{(a) The difference in monopole excitation energy between the scaling and constraint calculations ($DE=E_S-E_C$) with NL3 parameter set as a function of neutron number. (b) The resonance width is defined as $\Sigma=\frac{1}{2}\sqrt{E_s^2-E_C^2}$} for the isotopic series as a function of neutron number.
\label{fig7}
\end{figure}
After knowing that the present technique (RETF) is very efficient for the superheavy region, we calculate the GMR in the framework of RETF with the NL3 set for the $Z = 110-118$ nuclei. Both the scaling and constraint calculations are performed to evaluate the monopole excitation energy and the finite nuclear incompressibility. Then we look for any correlation among them, so that it will be useful for further studies. The superheavy nuclei lie on the stability line, but are very unstable due to excessive Coulomb repulsion. To neutralise the repulsive Coulomb effect, a large number of neutral neutrons are needed and are responsible for making a highly asymmetric finite nuclear system. As a result, the state dependence of the nuclear force plays an important role and makes the situation very complicated and very different from the normal $\beta-$stable nuclei. Since the considered nuclei are very asymmetric with a strong Coulomb force, these nuclei are vulnerable to both $\alpha-$ and $\beta-$decays. It is well understood that the nuclear force is highly state-dependent. The $singlet-single$ and $triplet-triplet$ nucleon-nucleon interaction is repulsive, whereas the $singlet-triplet$ is very attractive \cite{Patr1992,Patr2004,Kaur2020}. Due to the presence of a large number of neutrons in superheavy nuclei, the $singlet-singlet$ and $triplet-triplet$ parts of the nuclear force in the superheavy region play the most dominant role, and the stability of the nucleus becomes questionable.  In these circumstances, the giant monopole resonance (GMR) is a main source of information on the incompressibility of nuclei and nuclear matter~\cite{Blai1980}. 
Since the incompressibility of the finite nuclei has interesting relations with the nuclear matter incompressibility, the study of GMR for the superheavy nuclei is quite intriguing.

In Fig. \ref{fig6}, the scaling and constraint monopole excitation energies for Ds, Rg, Cn, Nh, Fl, Mc, Lv, Ts and Og are shown. As usual, in the known cases, here also $E_S$ is constantly overestimated by $E_C$. At N = 180, all the nuclei predict a similar scaling monopole excitation energy $E_S$ = 12.75 MeV. The $E_C$ is shown in panel (b) of Fig. \ref{fig6}.  In the constraint excitation energy, we also get the same results for N = 190, i.e., $E_C = 11.75$ MeV, similar to the $E_S$ with a slight shifting in the neutron number. The variation in excitation energy decreases with the higher value of atomic number $Z$ for both scaling and constraint calculations. The rate of change of scaling energy (0.028 MeV per nucleon) is less than the rate of change in constraint energy (0.031 MeV per nucleon) with an increase in neutron number. Throughout the superheavy region ($Z = 110-118$), the energy difference between the $E_S$ and $E_C$ for all the cases  ($DE=E_S-E_C$) is less than 1 MeV, suggesting a sharp spectrum in the monopole resonances as compared to the O and Zr isotopes. The $DE$ as a function of neutron number is shown in Fig. \ref{fig7}(a). It continues to increase slowly with N, and most isotopes meet near N = 210. On the other hand, if we analyse the resonance width defined as $\Sigma=\frac{1}{2}\sqrt{E_S^2-E_C^2}$ goes on decreasing with N and becomes minimum at about neutron number $N=290-220$ for various isotopic series (see Fig. \ref{fig7}(b)). It is extreme left for the lightest Ds series and extreme right for the heavy Og chain. The maximum width $\Sigma$ for the majority case is within $2.3-2,44$ MeV inferring a sharp resonance peak and indicating a perfect collective monopole resonance.

\begin{figure}
\includegraphics[width=1.0\columnwidth]{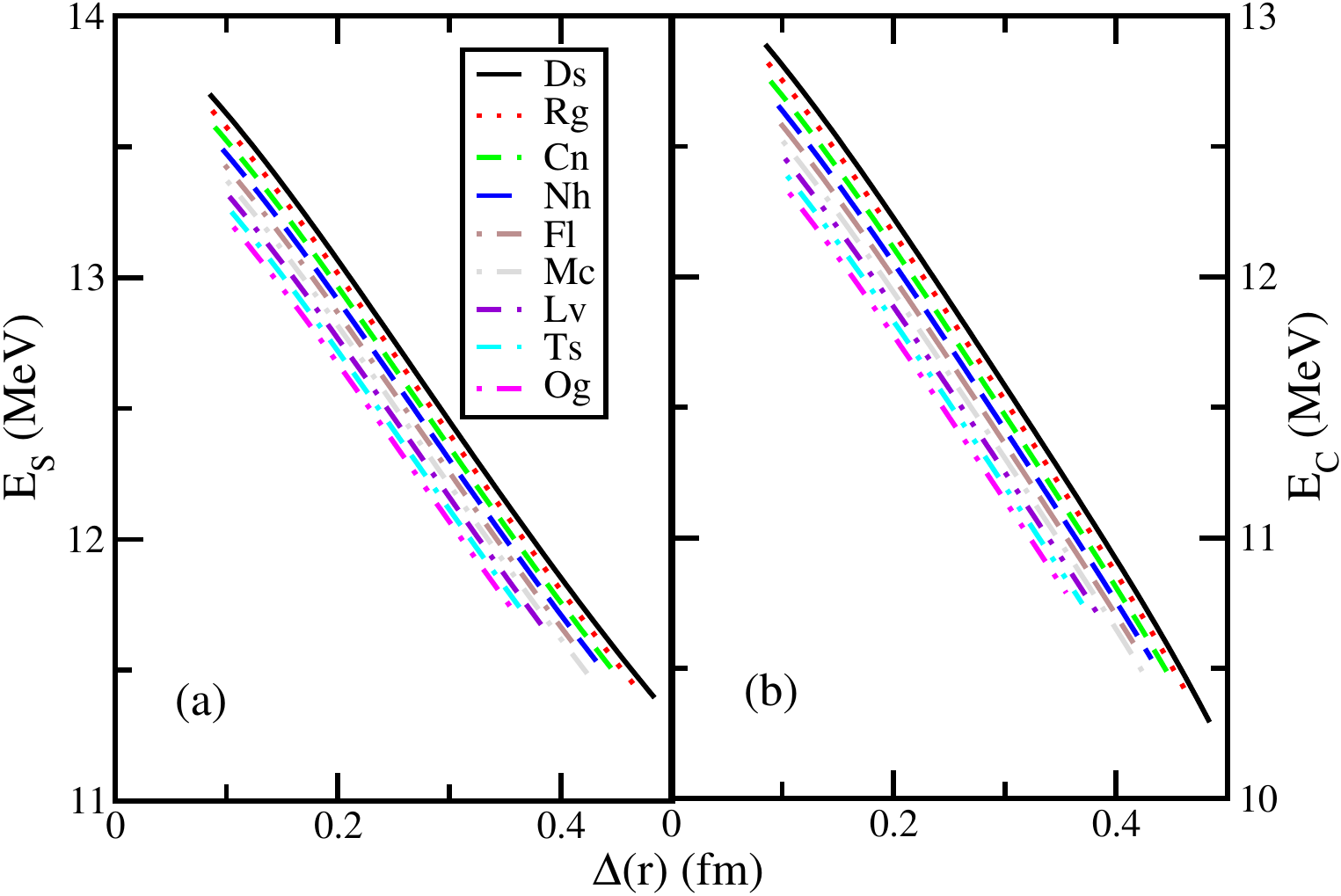}
\caption{The giant monopole resonance excitation energy for superheavy nuclei as a function of neutron-skin thickness $\triangle{r}=R_n-R_p$.}
\label{fig8}
\end{figure}
The relation of giant monopole resonance with the neutron-skin thickness can be seen in Fig. \ref{fig8}. The $E_S$ is given in Fig. \ref{fig8}(a) and $E_C$ is in Fig. \ref{fig8}(b). For both the $E_S$ and $E_C$, we get a negative correlation with the neutron-skin thickness $\triangle{r}=R_n-R_p$, i.e., both the excitation energies decrease with neutron number and neutron-skin thickness $\triangle{r}$ increases with N due to excess in neutron in an isotopic chain. We made a simple linear regression fitting for both the $E_S$ and $E_C$ separately, and the fitting equations are given as $E_S=14.044-5.9514\times{\triangle{r}}$ and $E_C=13.25-6.3228\times{\triangle{r}}$ with a correlation coefficient $\zeta=$99.9999 for both cases.

\subsection{The incompressibility of superheavy nuclei $Z=110-118$ }
\begin{figure}
\includegraphics[width=0.9\columnwidth]{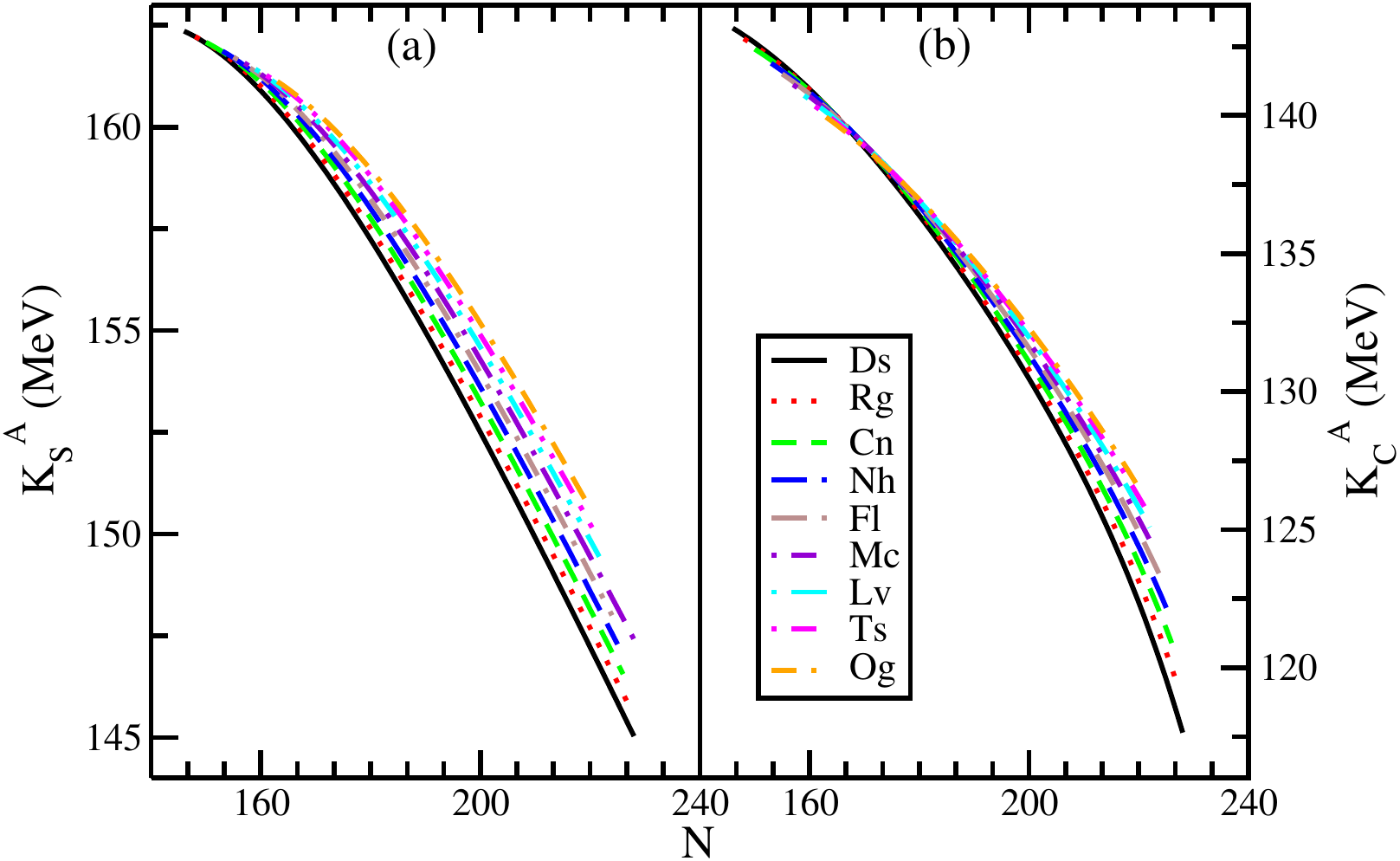}
\caption{The finite nuclear incompressibility $K^A$ for the superheavy nuclei for $Z=110-118$ as a function of neutron number in scaling and constraint methods.}
\label{fig9}
\end{figure}
Once we know the excitation energy $E_x$, from Eqn. (18), we can evaluate the incompressibility $K^A$ for a finite nucleus with mass number $A$. The constraint and scaling results of $K^A$ are shown in Fig. \ref{fig9} as a function of neutron number. It is clear from the results that when we increase the number of neutrons, the scaling and constraint incompressibility monotonously decrease. The relation of $K^A$ and N looks like a negative correlation as discussed in the earlier sub-section for the excitation energy $E_x$ with the neutron-skin thickness $\triangle{r}$. That means, the $K^A$ decreases with N with some proportionality relation. The rate of change of scaling incompressibility $K^A_{S}$ is more than that of $K^A_C$ with neutron number. Similar to the excitation energy, here also we get a common value of $K_C$  for all considered isotopes at $N = 170$, which is shifted to $N=157$ in $K_S$ calculations. The difference in $K^A$ between the scaling and constraint, $(\Delta K=K^A_S - K^A_C)$ is within  $19-28$ MeV.


\section{SUMMARY AND CONCLUSIONS} \label{summary}
In the present study, we calculate the giant monopole resonances for the heaviest known region of the finite nuclei mass table within the relativistic extended Thomas-Fermi formalism using the NL3 parameters set. Before going to compute the collective properties, like $E^x$, and $K^A$, we tested the model and parametrisation for properties of known nuclei in the light and superheavy regions. In this context, the binding energy is compared with the Hartree result and experimental data for a few known cases. We find that the relativistic extended Thomas-Fermi calculations are more suitable with a large number of nucleonic systems. That is why the present method is quite an accurate and time-saving formalism for the calculations of a wide range of isotopes.

Once we established the RETF formalism and the application of scaling and constraint methods to the analysis of collective properties of light and medium mass finite nuclei, we extended the calculations to the superheavy nuclei. We have shown that, unlike the light/medium mass isotopes,  the excitation energy and incompressibility for finite nuclei in the $Z=110-118$ region decrease monotonously in an isotopic chain. Since the scaling and constraint methods predict the higher and lower sides of the monopole spectrum, their difference estimates the width of the resonance. We get a sharper resonance, because both the scaling and constraint calculations are found to be almost similar with a smaller value of $DE$, i.e. $DE\sim 1$ MeV. Although the difference in $E_S$ and $E_C$ gives an idea about the resonance width, the actual width $\Sigma$ is defined as $\Sigma=\frac{1}{2}{\sqrt{E_S^2-E_C^2}}$. In this case, we get the maximum $\Sigma$ within the range $2.3-2.44$ MeV for the considered band of superheavy nuclei.  The maximum $\Sigma$ for a particular element also depends on the neutron number. Since the participation of the Coulomb effect plays an important role in the $\Sigma$, we get the maximum width for the lower value of N for lighter Z elements (here for Ds) and vice versa (here for Og). That means, in the Ds isotopic series, the maximum $\Sigma$ is at N = 190, and for the Og chain it is at N = 220. The incompressibility for finite nuclei $K^A$ is found to be similar for the excitation energy. That means, the rate of change of $K^A$ in an isotopic series with neutron number is the same as that of the excitation energy $E^x$ in the superheavy region.

\section*{Acknowledgments}
This work has been supported by the Science and Engineering Research Board (SERB), DST, India, under Ramanujan Fellowship File No. RJF/2022/000140.  


\end{document}